\documentclass[aip,apl, reprint, floatfix]{revtex4-1}

\usepackage{graphicx}
\usepackage{dcolumn}
\usepackage{bm}
\usepackage {color}

\begin{document}

\preprint{AIP/123-QED}

\title{Enhancement of the spin pumping efficiency by spin-wave mode selection}

\author{C.~W. Sandweg}
 \email{sandweg@physik.uni-kl.de}
\affiliation{%
Fachbereich Physik and Forschungszentrum OPTIMAS,
Technische Universit\"at Kaiserslautern, 67663 Kaiserslautern, Germany}%

\author{Y. Kajiwara}
\affiliation{
Institute for Materials Research, Tohoku University, Sendai 980-8577, Japan}
\author{K. Ando}
 \affiliation{
Institute for Materials Research, Tohoku University, Sendai 980-8577, Japan}
\author{E. Saitoh}
 \affiliation{
Institute for Materials Research, Tohoku University, Sendai 980-8577, Japan}
 \affiliation{Advanced Research Center, Japan Atomic Energy Agency, Tokai 319-1195, Japan}
 \affiliation{CREST, Japan Science and Technology Agency, Sanbancho, Tokyo 102-0075, Japan}
 \affiliation{PRESTO, Japan Science and Technology Agency, Sanbancho, Tokyo 102-0075, Japan}

\author{B.~Hillebrands}
 \affiliation{%
Fachbereich Physik and Forschungszentrum OPTIMAS,
Technische Universit\"at Kaiserslautern, 67663 Kaiserslautern, Germany}%

 \date{\today}

\begin{abstract}
The spin pumping efficiency of lateral standing spin wave modes in a rectangular Y$_\mathrm{3}$Fe$_\mathrm{5}$O$_\mathrm{12}$/Pt sample has been investigated by means of the inverse spin-Hall effect (ISHE). The standing spin waves drive spin pumping, the generation of spin currents from magnetization precession, into the Pt layer which is converted into a detectable voltage due to the ISHE. We discovered that the spin pumping efficiency is significantly higher for lateral standing surface spin waves rather than for volume spin wave modes.  The results suggest that the use of higher-mode surface spin waves allows for the fabrication of an efficient spin-current injector. \end{abstract}

\maketitle

The field of spintronics, a prospering class of efficient memories and computing devices based on electron spins, has become of great interest throughout the last decade. The aim of a future spintronics device is to overcome the limits of ordinary electronics devices by controlling the magnetization dynamics. A promising approach for propelling this new technology is the precise control as well as the manipulation of spin current.\cite{Ando 2008}

In this context, the spin pumping mechanism \cite{Tserkovnyak, Mizukami, Costache} is of immense importance for the generation of a spin current in paramagnetic materials since it is emitting a pure spin current at the interface between a ferromagnet or a ferrimagnet and a paramagnet. In contradiction to other methods generating a spin current, there is no electrical current driven through the interface in this case. This important property of the spin pumping opens the window to a completely new field of innovative electrical circuit designs and components, especially in combination with magnetic and paramagnetic metals as well as magnetic insulators.\cite{Kajiwara} The spin pumping mechanism can be observed directly via inverse spin-Hall effect (ISHE). \cite{Saitoh 2006, Valenzuela, Ando 2009, Kimura, Takahashi, Ando PRB, Ando 2009-2, Takeuchi} It induces an electromotive force in the paramagnetic metal which can be detected using a sensitive voltmeter. 

Mainly the pumping of the uniform precession mode was in the focus of previous investigations, but also the study of the spin wave spectrum and the corresponding spin pumping efficiencies is of crucial importance for a successful design of a future spintronics device involving spin pumping. 

In this letter we show that surface spin wave modes exhibit significantly greater efficiencies of spin pumping in Y$_\mathrm{3}$Fe$_\mathrm{5}$O$_\mathrm{12}$ rather than volume spin waves. These informations will be useful for developing spin-current injectors based on spin pumping.

\begin{figure}
\begin{center}
\scalebox{1}{\includegraphics[width=8.5 cm]{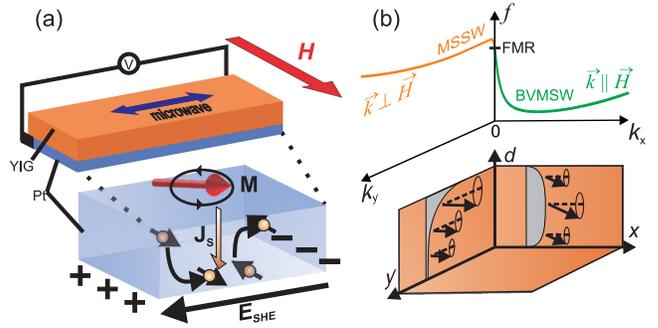}}
\end{center} 
\caption{\label{Fig1} (Color online) (a) Schematic illustration of the spin pumping and inverse spin-Hall effect in a YIG/Pt sample. The electromotive force is measured by attaching a nanovoltmeter to the Pt layer.
(b) Dispersion relation and distribution of the dynamic magnetization of magnetostatic surface modes (MSSW) and backward-volume magnetostatic waves (BVMSW) in YIG.}
\end{figure}

Figure~\ref{Fig1}~(a) shows the schematic illustration of the sample used in the present studies. The sample consists of a 2.1~$\mu$m thick (111)-single-crystal Y$_\mathrm{3}$Fe$_\mathrm{5}$O$_\mathrm{12}$ (YIG) film grown on a Gd$_\mathrm{3}$Ga$_\mathrm{5}$O$_\mathrm{12}$ single-crystal substrate by liquid phase epitaxy and a 10~nm thick Pt layer sputtered onto the YIG film . The sample has a rectangular shape with the width $x=1$~mm and the length $y=4$~mm. In addition, two electrodes are attached to the ends of the Pt layer and wired to a Keithley 2182A Nanovoltmeter (see Fig.~\ref{Fig1}~(a)) for a precise and stable measurement of the electromotive force $V$ generated by the ISHE. 

During the measurement the sample is placed in a JEOL JES-FA200 microwave absorption spectrometer so that it is in the center of a TE$_\mathrm{011}$ microwave cavity. Thus, the magnetic-field component of the mode is maximized and the electric field component is minimized respectively. A microwave mode with $f=9.441$~GHz is excited and sent to the cavity and in addition, a tunable external magnetic field $H$ is applied at the same time. In the presented measurement, the microwave magnetic field  and the bias magnetic field are aligned in-plane of the investigated sample.
The microwave absorption spectrometer uses a lock-in amplifier and therefore the derivative of the microwave absorption intensity $I$ with respect of the magnetic field $dI/dH$ is measured. When $H$ and $f$ fulfill the resonance condition of a magnetic mode, a spin current is resonantly injected into the Pt layer by the mechanism of spin pumping. Afterwards, the injected spin current is converted into a charge current by the ISHE and can be detected using the voltmeter. Thus, we measured simultaneously the microwave absorption signal and the electromotive force $V$ between the electrodes connected to the Pt layer.

Figure~\ref{Fig1}~(b) shows a sketch of the dispersion relation and the distribution of the dynamic magnetization of two spin waves for an in-plane magnetized YIG film.\cite{Kalinikos} At $k=0$ the mode of the uniform precession, the ferromagnetic resonance (FMR), is located. If the wavevector is non-zero, two different types of propagating spin waves can be distinguished. Magnetostatic surface spin waves (MSSW) have a wavevector which is oriented perpendicular to the external bias magnetic field and their distribution of the dynamic magnetization is strongly localized near the surface. In contrast, the wavevector of backward-volume magnetostatic spin waves (BVMSW) is oriented parallel to the bias field and their dynamic magnetization is distributed over the whole sample and small at the surface. In the present case, both types of spin waves are propagating in a rectangular YIG film of finite dimensions $x\times y$ and are reflected from the edges. If their wavevectors fulfill the conditions $k_{x}=n_{x}\pi/x$ and $k_{y}=n_{y}\pi/y$ with integers for $n_{x}$ and $n_{y}$, standing waves ($n_{y}$,$n_{x}$) are formed.\cite{Barak} They can be observed in the spin wave resonance (SWR) spectrum  for odd integers of ($n_{y}$,$n_{x}$) only since the net magnetic moment is zero for even numbers.

\begin{figure}
\begin{center}
\scalebox{1}{\includegraphics[width=8.5 cm]{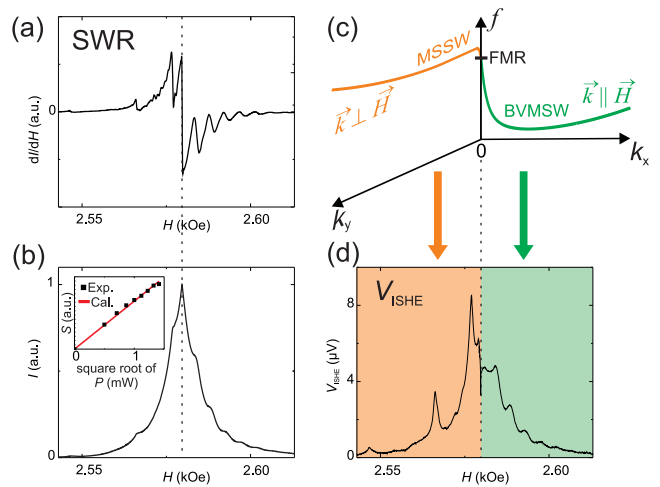}}
\end{center} 
\caption{\label{Fig2} (Color online) (a) Spin wave resonance (SWR) spectrum of the YIG/Pt sample at 2~mW. The dashed line shows the magnetic field position ($H$) of the ferromagnetic resonance (FMR). (b) Integrated SWR spectrum at 2~mW. $I$ denotes the microwave absorption intensity. In the inset, the integral intensity $S$ of the $I$-spectrum is drawn as a function of square root of the microwave power. (c) The schematic dispersion relation illustrates the positions of the MSSW and the BVMSW in the resonance spectrum  (d) Spectrum of the electromotive $V_\mathrm{ISHE}$ as a function of the bias magnetic field. The dashed line indicates the position of the FMR.}
\end{figure}

Figure~\ref{Fig2}~(a) shows the spin wave resonance spectrum of the YIG/Pt sample for a microwave power of 2~mW, a fixed frequency of 9.441~GHz and a magnetic field range from 254.3~mT to 261.3~mT. In the spectrum $dI/dH$ refers to the derivative of the microwave absorption intensity with respect to the magnetic field. Next to the main peak, which corresponds to the FMR, multiple resonance signals appear in the spectrum. These signals correspond to standing spin waves which exist due to the confinement of the investigated rectangular sample. 
In order to examine the microwave absorption intensity which has been absorbed by each mode, the SWR spectrum has been integrated, as shown in Fig.~\ref{Fig2}~(b). In the inset the integral intensity $S$ of the $I$-spectrum is drawn as a function of the square root of the microwave power. The linear slope proves that the microwave absorption is not saturated and therefore nonlinear effects will not influence the measurement. \cite{Schreurs}

Using the dispersion relations for MSSW and BVMSW it is possible to identify the resonance signals to the left and to the right of the FMR signal. As shown in Fig.~\ref{Fig2}~(c), MSSW have frequencies above the FMR and BVMSW below the FMR. By changing the magnetic field at a fixed frequency, the whole dispersion relation is shifted up or down. In this way, magnetic resonances lower than the FMR position corresponds to MSSW modes and higher field positions to BVMSW modes in accordance with the dispersion relation. The electromotive force $V_\mathrm{ISHE}$ measured with the nanovoltmeter is presented in Fig.~\ref{Fig2}~(d). The dashed line indicates the peak position of the FMR. To the left of this line the resonant electromotive force signals from the MSSW appear and on the right, the signals from the standing BVMSW modes are located.   

\begin{figure}
\begin{center}
\scalebox{1}{\includegraphics[width=6 cm]{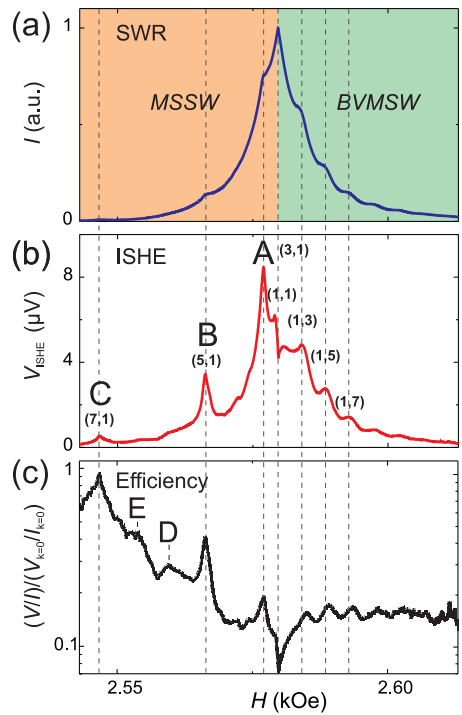}}
\end{center} 
\caption{\label{Fig3} (Color online) Comparison between the SWR and the V$_\mathrm{ISHE}$ spectrum. The V$_\mathrm{ISHE}$ spectrum shows resonant signals from MSSW modes ($n_{y}$,$n_{x}$)~=~(3,1), (5,1) and (7,1), which are hardly observable in the SWR spectrum. On the other hand, the BVMSW modes (1,3), (1,5) and (1,7) are visible in both spectra. The figure below shows the spin pumping efficiency which confirms this observation.}
\end{figure}

The main aim of the presented work is to evaluate the spin pumping efficiency for the different lateral standing spin wave modes. The spectrum of the electromotive force shows three peaks due to standing MSSW modes with ($n_{y}$,$n_{x}$)~=~(3,1), (5,1) and (7,1) (see ${\bf A}$, ${\bf B}$, ${\bf C}$ in Fig.~\ref{Fig3}~(b)), which are significantly pronounced but barely detectable in the SWR spectrum, as presented in Fig.~\ref{Fig3}~(a). The standing modes (1,3), (1,5) and (1,7) of the BVMSW are also clearly observable in the ISHE spectrum but not with such a significance as in the MSSW case. This indicates that the spin pumping efficiency is greater for MSSW standing modes as for BVMSW standing modes. 

In order to analyze the different spin pumping efficiencies in a more quantitative manner, the resonant electromotive force spectrum $V_\mathrm{ISHE}$ is divided by the integral intensity of the SWR spectrum $I$, which reflects the microwave absorption intensity by each spin wave mode. Figure~\ref{Fig3}~(c) shows the obtained spin pumping efficiency $V_\mathrm{ISHE}/I$ in logarithmic scale. The efficiency of the standing BVMSW modes remains constant, nearly independent of higher mode numbers. This demonstrates that the spin pumping efficiency of volume modes is not sensitive to the mode numbers in accordance with previous results.\cite{Ando 2009} On the other hand, the efficiency $V_\mathrm{ISHE}/I$ is clearly enhanced for MSSW modes (see the left branch of Fig.~\ref{Fig3}~(c)). The efficiency curve also clarifies that the spin pumping efficiency increases with higher mode numbers of the MSSW. 

This behavior can be attributed to different distributions of the dynamic magnetization over the sample thickness for the different standing spin waves. MSSW are strongly located at the surface and the dynamic magnetization decays exponentially. Thus, the coupling of the conduction electrons with the spin wave modes is enhanced. In contrast, the BVMSW are volume waves with a dynamic magnetization distributed over the whole sample and therefore the coupling interface is smaller than for surface waves. Consequently, the ratio between the electromotive force signal and the microwave absorption intensity is even enhanced for higher MSSW modes. Owing to this enhanced sensitivity for the surface mode detection, we recognize two additional peaks in the $V_\mathrm{ISHE}$ spectrum, i.e. peak ${\bf D}$ and ${\bf E}$ in Fig.~\ref{Fig3}~(c), which are not detectable in the conventional microwave absorption measurement (see Fig.~\ref{Fig2}~(a) and Fig.~\ref{Fig3}~(a)). They are likely modes with both ($n_{y}$,$n_{x}$)~$>$~1. This suggests that the ISHE method may provide a powerful tool for surface spectroscopy of spin dynamics. 

In summary, we demonstrated that standing surface spin waves have a extensively higher spin pumping efficiency rather than backward-volume waves. Thus, the efficiency of a spin-current generator can be increased by selecting the spin wave modes.

The authors thank A.~A.~Serga and A.~V.~Chumak for valuable discussions. This work was supported by the DFG within the SFB/ Transregio 49 'Condensed Matter Systems with Variable Many-Body Interactions', by a Grant-in-Aid for Scientific Research Priority Area 'Creation and control of spin current' (19048009,19048028), a Grant-in-Aid for Scientific Research (A), the global COE for the 'Materials integration international centre of education and research' all from MEXT, Japan, and a Grant for Industrial Technology Research from NEDO, Japan.

\end{document}